\documentclass{sf2a-conf2019}
\usepackage{graphicx}
\usepackage{hyperref}
\usepackage[]{natbib}  
\usepackage{epstopdf}

\def\BibTeX{{\rm B\kern-.05em{\sc i\kern-.025em b}\kern-.08em
    T\kern-.1667em\lower.7ex\hbox{E}\kern-.125emX}}
\bibpunct{(}{)}{;}{a}{}{,}  

\begin{document}

\TitreGlobal{SF2A 2019}




\title{Two examples of how to use observations of terrestrial 
planets orbiting in temperate orbits around low mass stars to 
test key concepts of planetary habitability.}



\runningtitle{Observational tests of key concepts of planetary habitability.}

\author{M. Turbet}\address{Observatoire astronomique de l\'\  Universite de Geneve, 51 chemin des Maillettes, 1290 Sauverny, Switzerland}

\setcounter{page}{237}


\maketitle


\begin{abstract}
Terrestrial planets in temperate orbit around very low mass stars are 
likely to have evolved in a very different way than solar system planets, and in particular Earth. 
However, because these are the first planets that are and will be accessible
for in-depth atmosphere, clouds and surface characterizations with existing and forthcoming telescopes, 
we need to develop the best possible observational strategies to 
maximize the scientific return from these characterizations.
Here I discuss and expand on the recent works of \citet{Bean:2017} and \citet{Turbet:2019aa} to show that 
terrestrial planets orbiting in temperate orbits around very low mass stars are potentially an excellent 
sample of planets to test how universal the 
processes thought to control the habitability of solar system planets and in particular Earth are.
Precise measurements of density or atmospheric CO$_2$ concentration for planets located both inside and outside the 
Habitable Zone could be used to statistically test habitability concepts such as the silicate-weathering feedback, CO$_2$ condensation, or 
runaway greenhouse, which have been identified as key processes controlling the present and past habitability of Venus, Mars and Earth.
\end{abstract}




\section{Introduction}

As of August 2019, astronomers have already detected about forty exoplanets in temperate orbit 
\citep{Pepe:2011,Tuomi:2013,Borucki:2013,Anglada:2013,Quintana:2014,Lissauer:2014,Anglada:2014,Torres:2015,Crossfield:2015,Wright:2016,Gillon:2016,Morton:2016,Anglada:2016,Crossfield:2016,Gillon:2017,Luger:2017,Astudillo-Defru:2017,Bonfils:2018,Diaz:2019,Tuomi:2019,Zechmeister:2019}, 
with masses or radii or sometimes even both that are similar to the Earth. 
Most of these recently detected exoplanets are orbiting around nearby, very low mass stars. 
This specificity make them not only easier to detect, but also easier to characterize with 
respect to planets orbiting more massive, e.g. solar-type stars. In-depth characterization of these exoplanets could be achieved through:
\begin{enumerate}
 \item combined mass and radius precise measurements. This allows to estimate the planet density, 
and thus to gain information on its bulk interior and possibly atmospheric composition.
 \item atmospheric, clouds and/or surface measurements, through a variety of techniques such as transit 
spectroscopy, direct imaging, secondary eclipse or thermal phase curves. 
\end{enumerate}

\medskip

However, planets orbiting around very low mass stars have at least two characteristics that are likely to make them 
evolve very differently from solar system planets, and in particular Earth. These two characteristics are:
\begin{enumerate}
 \item \textbf{A hot history.} Very low mass stars can stay for hundreds of millions of years in the Pre Main Sequence (PMS) phase, 
a phase during which their luminosity can decrease possibly by several 
orders of magnitude \citep{Chabrier:1997,Baraffe:1998,Baraffe:2015}. During this PMS phase, 
planets are exposed to strong irradiation, which make them really sensitive to atmospheric processes such 
as runaway greenhouse \citep{Ramirez:2014c}, indicating that 
all the so-called volatile species (e.g. H$_2$O, CO$_2$, CH$_4$, NH$_3$) and most of their byproducts 
must be in gaseous form in the atmosphere. Note that the runaway greenhouse atmospheric process is discussed in more details below.
 \item \textbf{An exposition to strong atmospheric escape.} Very low mass stars emit much more high energy X/EUV photons than solar-type stars, 
in proportion to their total bolometric emission \citep{Ribas:2017}, exposing therefore the atmosphere of close-in planets to 
strong atmospheric erosion mechanisms such as hydrodynamic escape \citep{Lammer:2009,Zahnle:2017,Bolmont:2017}. 
\end{enumerate}

\medskip

Combining these two previous constraints with a numerical planet population synthesis model 
lead \citet{Tian:2015} to infer that terrestrial planets in temperate orbit around 
very low mass stars are likely to end up in two very different states: (i) If the initial amount of 
volatile species present at the time of the planet's formation exceed what can be lost through atmospheric erosion processes, then 
the planet should remain volatile-rich, and likely water-rich since water 
is the most abundant volatile species, and also the most likely to condense on the surface among all common volatile species.
(ii) Otherwise, the planet would have to be completely dry by the end of the PMS phase, 
but could later have been replenished with some volcanic and/or volatile gases delivered by impacts.
In summary, these planets are likely to be 
either (i) \textbf{extremely water-rich} or (ii) \textbf{water-poor}\footnote{The Earth and other solar system 
terrestrial planets fit in this second category.}, 
i.e. planets that have low enough water to have continents present.


\medskip


\medskip

Despite exotic characteristics, here I arguee that planets orbiting very 
low mass stars are still potentially an excellent sample of planets 
to test processes thought to control the past and present habitability 
of solar system planets, and therefore an excellent way to test how universal these processes are.
In particular, I discuss and expand on two processes that are thought to be key of the Earth's habitability, 
and which have led to proposals of observational tests \citep{Bean:2017,Turbet:2019aa} in extrasolar planet populations, 
namely (i) \textbf{the carbonate-silicate cycle}, that could be tested for the \textbf{water-poor} category of planets, and 
(ii) \textbf{the runaway greenhouse}, that could be tested for the \textbf{extremely water-rich} category of planets.

\section{First example: Testing the carbonate-silicate cycle and more broadly the CO$_2$ cycle}


\begin{figure}[ht!]
 \centering
 \includegraphics[width=0.8\textwidth,clip]{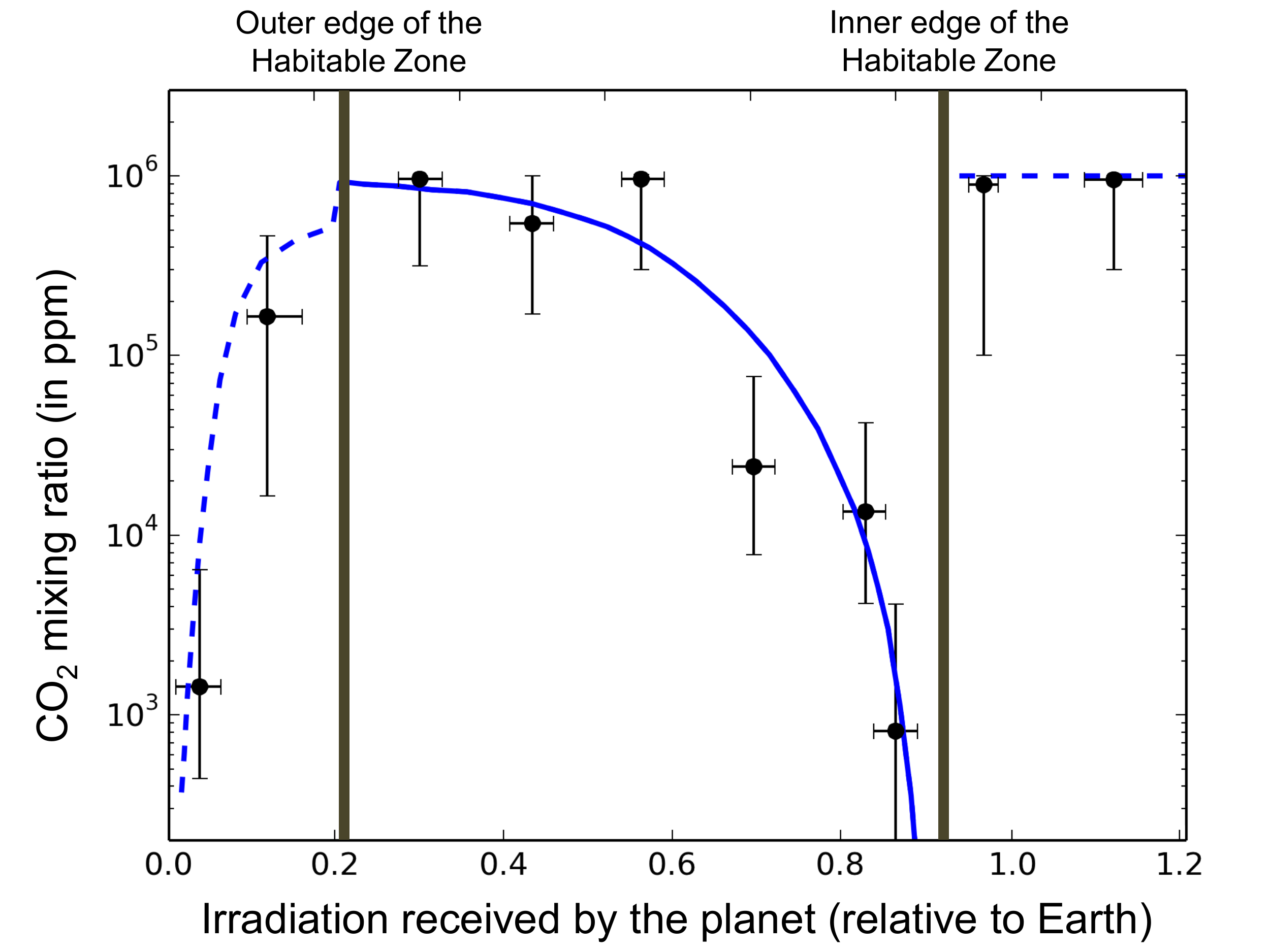}      
  \caption{This plot shows how measurements of CO$_2$ atmospheric mixing ratio for a sample of terrestrial-size planets 
spanning a wide range of irradiations could be used to statistically infer the 
existence of a CO$_2$ cycle, and even possibly the existence of a carbonate-silicate cycle. 
Between the inner and outer edges of the Habitable Zone \citep{Kopparapu:2013}, 
the blue solid curve (adapted from \citealt{Bean:2017}) shows the predicted CO$_2$ 
needed to maintain a surface temperature of 290~K. While planets located beyond the inner edge of the Habitable Zone 
are expected to accumulate large amount of CO$_2$, planets located below the outer edge of the Habitable Zone are expected 
to be depleted in CO$_2$ because of CO$_2$ condensation.
The black points are binned data for hypothetical planets.}
  \label{turbet:fig1}
\end{figure}

The CO$_2$ cycle is thought to be a key element for the stabilization of Earth's climate 
on geologically long timescale, through the carbonate-silicate cycle \citep{Walker:1981} which 
acts as a geophysical thermostat.  
This stabilizing cycle is thought to regulate the atmospheric CO$_2$ level in order to maintain surface temperatures 
that allow surface liquid water, based on two distinct processes: 
CO$_2$ degassing by volcanoes and silicate weathering, which strongly depends on temperature.
If a planet -- on which the carbonate-silicate feedback operates -- gets too warm, then 
the silicate weathering rate increases, which decreases 
the amount of CO$_2$ in the atmosphere, which further decreases the surface temperature of the planet. 
If a planet gets too cold, the silicate weathering rate decreases, and CO$_2$ accumulates through volcanic outgassing, which 
leads to surface warming through CO$_2$ greenhouse effect.

Assuming the carbonate-silicate cycle is a universal geochemical process, \citet{Bean:2017} proposed that it 
could be detected if we observe -- with statistical significance -- that Habitable Zone planets 
have a CO$_2$ content (or mixing ratio, for the observational point of view) that decreases with incident irradiation, 
as illustrated in Figure~\ref{turbet:fig1} (solid blue line). 
This is in fact what would be expected for planets that are sufficiently 
water-poor (first category) that they have both liquid water oceans or lakes, and continents. 
For planets very rich in water (second category), the CO$_2$ mixing ratio versus irradiation might look 
very different, because the CO$_2$ content is governed by other processes such as seafloor weathering 
or CO$_2$ oceanic dissolution \citep{Kitzmann:2015,Nakayama:2019}. 
I encourage future studies to better estimate how the CO$_2$ versus 
irradiation curve is expected to look like in the population of 
water-rich terrestrial planets.


In the water-poor limit planet population (i.e. planets that have oceans or lakes, and continents), 
I expand here the work of \citet{Bean:2017} on how the CO$_2$ content should vary as a function of irradiation 
for planets located outside the limits of the Habitable Zone:
\begin{enumerate}
 \item For planets receiving more irradiation than the inner edge of the Habitable Zone, water 
is expected to have completely evaporated into the atmosphere and 
thus to be exposed to photodissociation and subsequent atmospheric escape processes.  
This is likely what happened to Venus (see the introduction section of \citealt{Way:2016} 
for a recent review). 
Not only could the O$_2$ remaining in the atmosphere have oxidized the surface, thus producing CO$_2$; 
but also the absence of a hydrological cycle should have shut-down the silicate-weathering feedback, 
thus leading to the accumulation of CO$_2$ by volcanic degassing. Therefore, the CO$_2$ mixing ratio 
could reach unity for planets beyond the inner edge of the Habitable Zone (see dotted blue line in 
Figure~\ref{turbet:fig1}, right of the Inner edge of the
Habitable Zone).
 \item For planets receiving less irradiation than the outer edge of the Habitable Zone, 
CO$_2$ is limited by surface condensation \citep{Turbet:2017epsl,Turbet:2018aa}, 
which should be more and more severe as the planet is further out of the host star. Therefore, it is expected that 
for planets receiving less irradiation than the outer edge of the Habitable Zone, the CO$_2$ atmospheric mixing 
ratio should decrease with decreasing irradiation, 
with possibly a gap at the exact position of the outer edge of the Habitable Zone, 
due to the ice albedo feedback (see dotted blue line in 
Figure~\ref{turbet:fig1}, left of the Outer edge of the
Habitable Zone). However, this gap is likely to be small for planets orbiting very low mass stars because 
the ice albedo feedback should not be very effective, due to 
(i) the spectral properties of water ice and snow\citep{Joshi:2012,Shields:2013} 
and (ii) the fact that these planets are likely in synchronous rotation, with 
all ice trapped on the nightside \citep{Menou:2013,Leconte:2013aa,Turbet:2016}. 
\end{enumerate}


\medskip

CO$_2$ measurements could be attempted first through the transmission spectroscopy technique as soon as the James Webb Space 
Telescope (JWST) is operational, possibly through the 4.3~microns CO$_2$ $\nu_3$ absorption band, which has been shown 
to be one of the most accessible molecular absorption band in terrestrial-type 
atmospheres \citep{Morley:2017,Lincowski:2018,Fauchez:2018,Wunderlich:2019,Lustig-Yaeger:2019}. Not only this feature 
is present for a wide range of CO$_2$ mixing ratio, but it is also weakly affected by the presence 
of clouds and photochemical hazes (Fauchez et al., submitted to the Astrophysical Journal).

\section{Second example: Testing the runaway-greenhouse}

Planets similar to Earth but slightly more 
irradiated are expected to experience a runaway greenhouse transition, 
a state in which a net positive feedback between surface temperature, evaporation, and atmospheric opacity 
causes a runaway warming \citep{Ingersoll:1969,Goldblatt:2012}. 
This runaway greenhouse positive feedback ceases only when oceans have completely boiled away, 
forming an optically thick H$_2$O-dominated atmosphere. 
Venus may have experienced a runaway greenhouse transition in the past \citep{Rasool:1970,Kasting:1984}, and we expect that  
Earth will in $\sim$~600~million years as solar luminosity increases by $\sim$~6$\%$ compared to its present-day value \citep{Gough:1981}. 
However, the exact limit at which this 
extreme, rapid climate transition from a temperate climate (with most water condensed on the surface) to 
a post-runaway greenhouse climate (with all water in the atmosphere) 
would occur, and whether or not 
a CO$_2$ atmospheric level increase would affect that limit, 
is still a highly debated topic \citep{Leconte:2013nat, Goldblatt:2013, Ramirez:2014b, Popp:2016}. 
This runaway greenhouse limit is traditionally used to define the inner edge of the habitable zone \citep{Kasting:1993,Kopparapu:2013}.

Assuming the runaway greenhouse feedback is a universal atmospheric physics process, 
\citet{Turbet:2019aa} recently proposed that the runaway greenhouse could be identified through a radius gap at the 
position of the runaway greenhouse irradiation. \citet{Turbet:2019aa} actually showed 
two same planets -- but one being located just below and the other just above the runaway greenhouse 
irradiation threshold -- should have a different transit radius 
and which should be all the more different as the planet water content get higher. This radius difference or gap is 
a consequence of the runaway greenhouse radius inflation effect introduced in \citet{Turbet:2019aa}, resulting 
from the fact that for a fixed water-to-rock mass ratio, a planet 
endowed with a steam H$_2$O-dominated atmosphere has a much larger physical size than if all the water 
is in condensed form (liquid or solid). For Earth, the net radius increase should be around $\sim$~500~km, but is 
expected to be significantly larger (up to thousands of km) for planets with much larger total water content \citep{Turbet:2019aa}.

\begin{figure}[ht!]
 \centering
 \includegraphics[width=0.8\textwidth,clip]{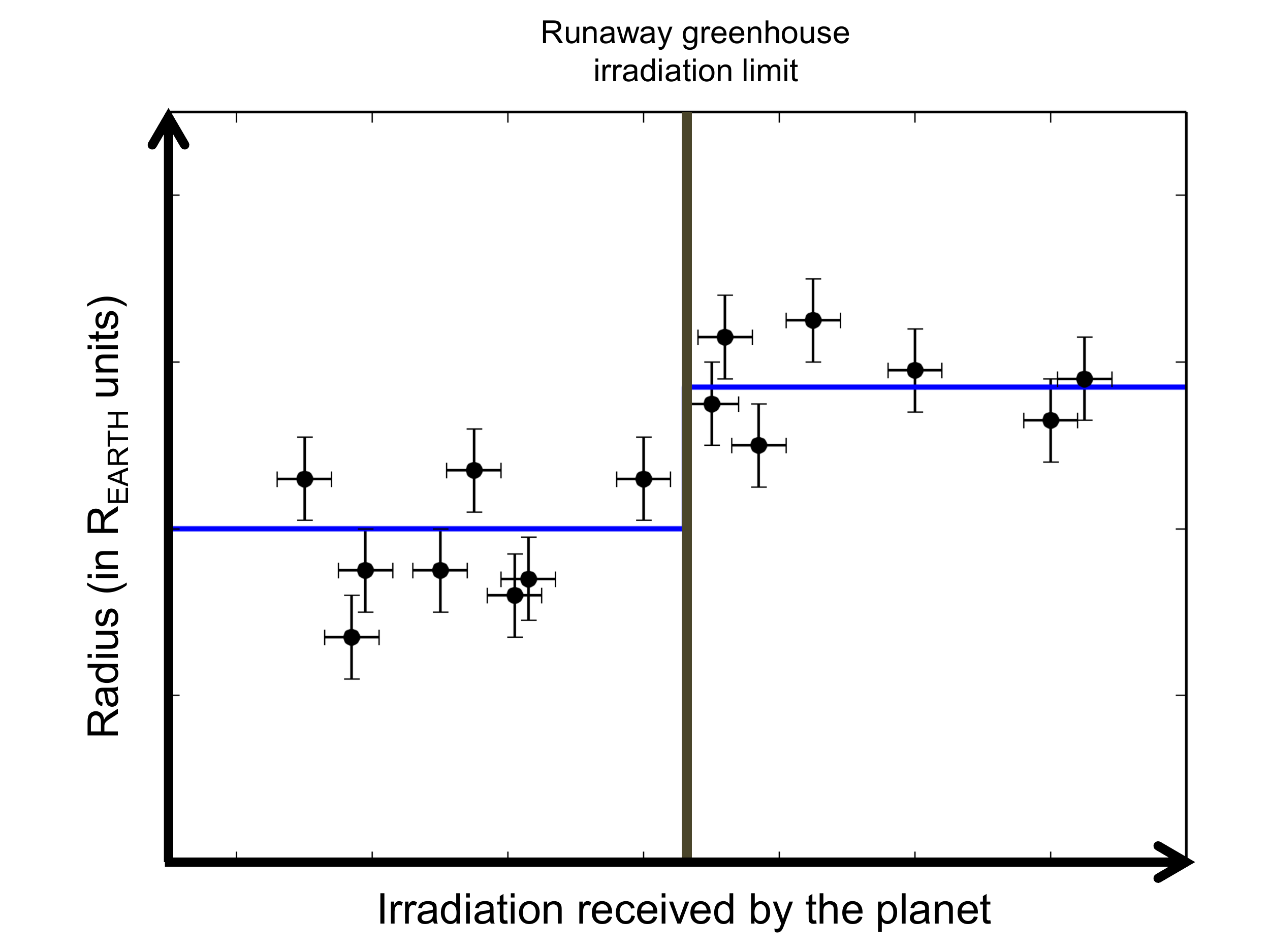}      
  \caption{This plot shows how combined measurements of masses and radii for a sample of terrestrial-size planets spanning 
irradiation on both sides of the runaway greenhouse irradiation limit could be used to validate the concept of runaway greenhouse. 
The blue solid curve (adapted from \citealt{Turbet:2019aa}) shows the predicted radius gap 
arising from the runaway greenhouse radius inflation effect. 
The black points are binned data for hypothetical planets which are in a fixed terrestrial mass range.}
  \label{turbet:fig2}
\end{figure}

\medskip

As a result, for a sample of water-rich planets\footnote{This runaway greenhouse radius inflation effect is expected to be absent 
from the population of water-poor planets orbiting low mass stars.}, 
the runaway greenhouse irradiation could be determined if 
we observe -- with statistical significance -- that planets located beyond the inner edge of the Habitable Zone 
have -- for a fixed terrestrial mass range -- a larger radius than planets located inside the Habitable Zone or colder, 
as illustrated in Figure~\ref{turbet:fig2} (solid blue line). 

Precise density measurements for terrestrial-size planets could be attempted by combining 
precise transit photometry with ongoing and upcoming space missions such as HST, TESS, CHEOPS and PLATO, with 
precise radial velocity mass measurements with ground-based spectrographs such as ESPRESSO, CARMENES or
SPIRou.

\section{Conclusions}

In this proceeding, I discussed and expanded on two possible observational 
strategies recently introduced in \citet{Bean:2017} and \citet{Turbet:2019aa} to constrain 
two key processes that are believed to be crucial to sustain habitability of solar system planets: the CO$_2$ cycle 
and the runaway greenhouse.
While the former could be first attempted as soon as JWST 
will be operational, the later could be tested with ongoing and future precise combined mass and radius measurements of 
terrestrial exoplanets.

\medskip

Although these strategies require more work \citep{Checlair:2019} to better constrain how to carry them out 
(how to make these observations? with which instruments? to what precision? what is the minimum number of planets needed? 
what are the best planets to be selected? how to deal with confounding factors? etc.), 
they demonstrate at least in theory how we could use observations and characterizations of exotic planets orbiting very low mass 
stars to test the universality of the processes that shape the habitability of planets in the solar system and 
possibly in many other exoplanetary systems.



\begin{acknowledgements}
This project has received funding from the European Union’s Horizon 2020 research and 
innovation program under the Marie Sklodowska-Curie Grant Agreement No. 832738/ESCAPE. 
M.T. would like to thank the Gruber Foundation for its generous support to this research.
\end{acknowledgements}

\bibliographystyle{aa}  
\bibliography{Turbet_S10} 

\end{document}